\documentclass[11pt]{article}
\usepackage{moriond,epsfig}

\def\Journal#1#2#3#4{{#1} {\bf #2}, #3 (#4)}
\def\Preprint#1#2{{\em astro-ph/#1}, ({#2})}

\def\APJ{{\em ApJ}}
\def\APJL{{\em ApJL}}
\def\APJSS{{\em ApJSS}}
\def\MN{{\em MNRAS}}

\def\be{\begin{equation}}
\def\ee{\end{equation}}
\def\bea{\begin{eqnarray}}
\def\eea{\end{eqnarray}}

\newcommand{\mt}[1]{\mbox{$\mathbf{#1}$}}
\newcommand{\VEV}[1]{\langle#1\rangle}

\begin{document}
\vspace*{4cm}
\title{Testing the isotropy of the Cosmic Microwave Backround}

\author{F. K. Hansen}

\address{Dipartimento di Fisica, Universit\`a di Roma `Tor Vergata', Via della Ricerca Scientifica 1, I-00133 Roma, Italy}

\maketitle\abstracts{I review the findings of an asymmetric distribution of large scale power in the \emph{WMAP} data and compare this to detections of non-Gaussianity on a part of the sky using local curvature properties.}

\section{Introduction}

The \emph{FIRAS} experiment on the \emph{COBE} satellite \cite{cobe1,cobe2} has shown that the temperature of the Cosmic Microwave Background (CMB) is isotropic to a very high degree. However, the angular resolution of the experiment was too low to make any significant statements on the isotropy of the spatial distribution of the tiny fluctuations in the CMB temperature. According to the cosmological principle of isotropy, the statistical properties of the CMB fluctuations should be the same in all directions on the sky. With the recent data from the \emph{WMAP} satellite \cite{WMAP} this has now been tested. Here, I will review the results of two different approaches to test the isotropy of the CMB using the \emph{WMAP} data and make a comparison between the results.

The angular power spectrum of the CMB fluctuations has been estimated on discs of various sizes centred in different directions\cite{p3}. It was found that for scales $\ell>40$ (corresponding to an angular scale of about $3^\circ$), the distribution of the CMB fluctuations is consistent with the hypothesis of isotropy, except for some signatures of possible foreground contamination around the first peak $\ell\sim220$. For the lower multipoles however, a strong difference between the north and the south (galactic and ecliptic) was found. This also leads to a difference in the estimate of the cosmological parameters between these hemispheres \cite{p5}. A similar large scale asymmetry/non-Gaussianity of the CMB fluctuation field has also been noted by other authors \cite{park,eriksen1,vielva,eriksen2,copi,curvat,wandelt,cruz,land} using different methods.

Using a test of non-Gaussianity based on the local curvature properties of a CMB field, a non-Gaussian signal has been found in one part of the sky \cite{curvat}. For a Gaussian field on the sphere, the local curvature properties are known \cite{dore} and a non-Gaussian signal may be revealed through tests of the curvature. Some non-Gaussianity tests have found signatures of residual foregrounds \cite{nas1,nas2,nas3,nas4}, but there have been other detections of non-Gaussianity on the whole or parts of the sky for which no easy explanation has been found (see ref. above). Here I will make a comparison between the asymmetric structure seen in the curvature and in the angular power spectrum and discuss whether there may be a common origin.

\section{The local power spectrum}

The power spectrum was estimated on differently
orientated hemispheres in the \emph{WMAP} data \cite{p3} using the Gabor transform formalism introduced by \cite{hansen1,hansen2} and extended in \cite{p3}. A brief outline of the method is provided here.

A Gaussian likelihood ansatz is adopted with the pseudo power
spectrum coefficients $\tilde C_\ell$ (the power spectrum obtained by making a spherical harmonic transform of the noisy CMB data on the cut sky) as input.  The likelihood can be written as
\begin{equation}
\label{eq:multilik}
\mathcal{L}=\frac{\mathrm{e}^{-\frac{1}{2}{\mathbf{d}}^\mathrm{T} {\mt{M}}^{-1}{\mathbf{d}}}}{\sqrt{2\pi \det{\mt{M}}}},
\end{equation}
where the elements of the data vector $\mathbf{ d}$ are given by the difference between the observed $\tilde{C}_\ell$ and its ensemble average, $d_i=\tilde{C}_\ell-\VEV{\tilde{C}_\ell}$. The correlation matrix is given by $M_{ij}=\VEV{d_id_j}$. Both of these depend on the full sky power spectrum $C_\ell$ which can be estimated by maximising this likelihood with respect to $C_\ell$.
The exact form of the matrix $\mt{M}$,
its dependence on the full sky power spectrum $C_\ell$ and
details of their computation are discussed in
\cite{p3,hansen1,hansen2}.

The above procedure was applied to the \emph{WMAP} data\footnote{These can be obtained at the LAMBDA website: \emph{http://lambda.gsfc.nasa.gov/}} applying the Kp2 galaxy and point source mask. The spectrum was estimated on 82 opposite pairs of hemispheres with their axis pointing in 82 different directions. The same procedure was carried out on 2048 simulated maps having the same beam and noise properties as the \emph{WMAP} map. For the multipole range $\ell=5-40$, the asymmetry was found to be strongest in the direction $(80^\circ,57^\circ)$ (galactic co-latitude and longitude) and only $0.7\%$ of the simulated maps had such a high maximum asymmetry (when comparing to the axis of maximum asymmetry for each simulation). For the lower multipoles ($\ell=5-20$), the axis of maximum asymmetry was closer to the galactic north pole. In figure \ref{fig:maps} I show the ratio of the power spectrum for the range $\ell=5-20$ between the 82 pairs of hemispheres. In the figure,  the centre of each hemisphere is indicated by a disc.

\section{The local curvature}

\begin{figure}
\psfig{figure=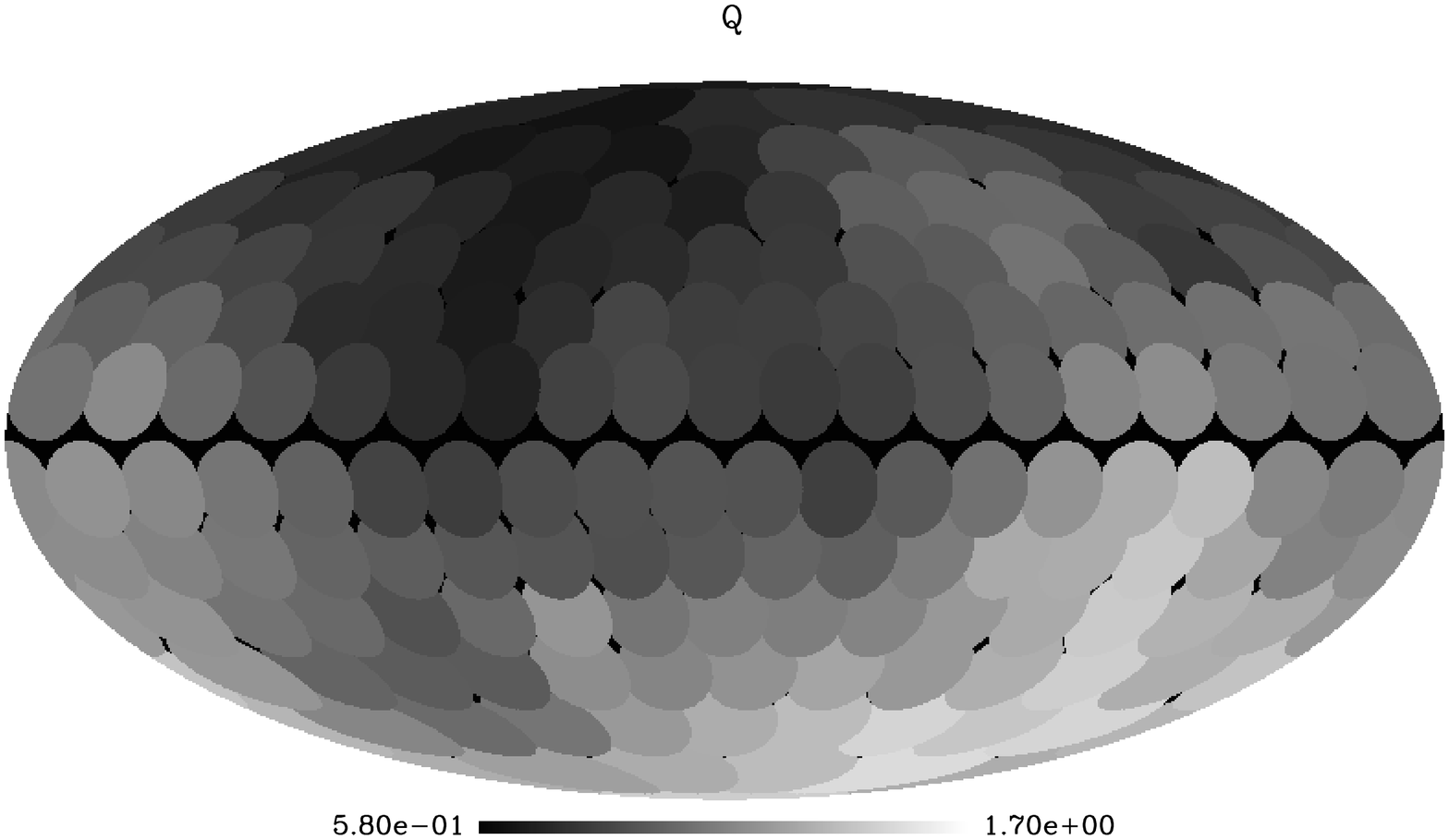,angle=90,height=3.cm,width=6.cm}\hspace*{2cm}
\psfig{figure=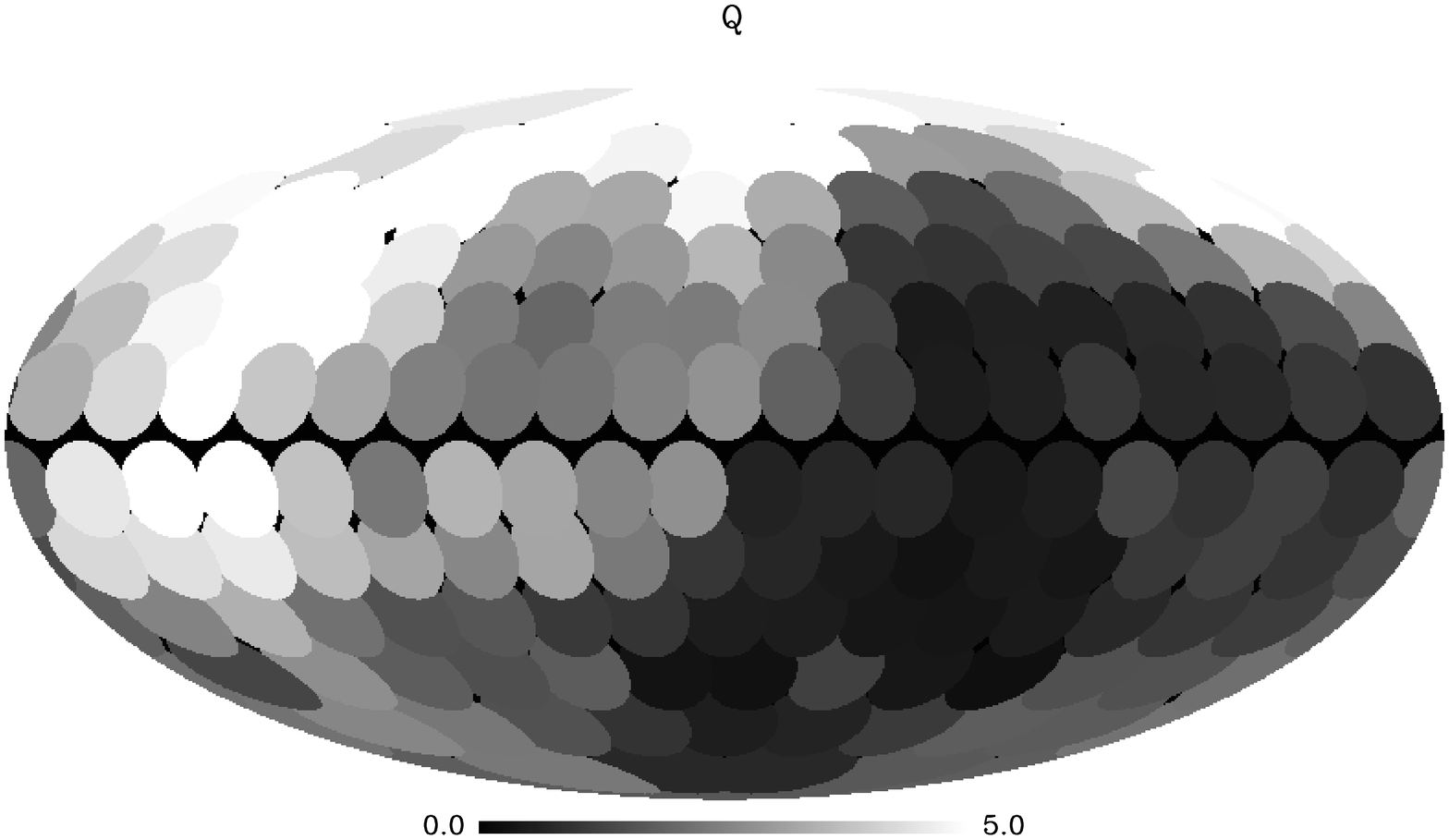,angle=90,height=3.cm,width=6.cm}\\
\psfig{figure=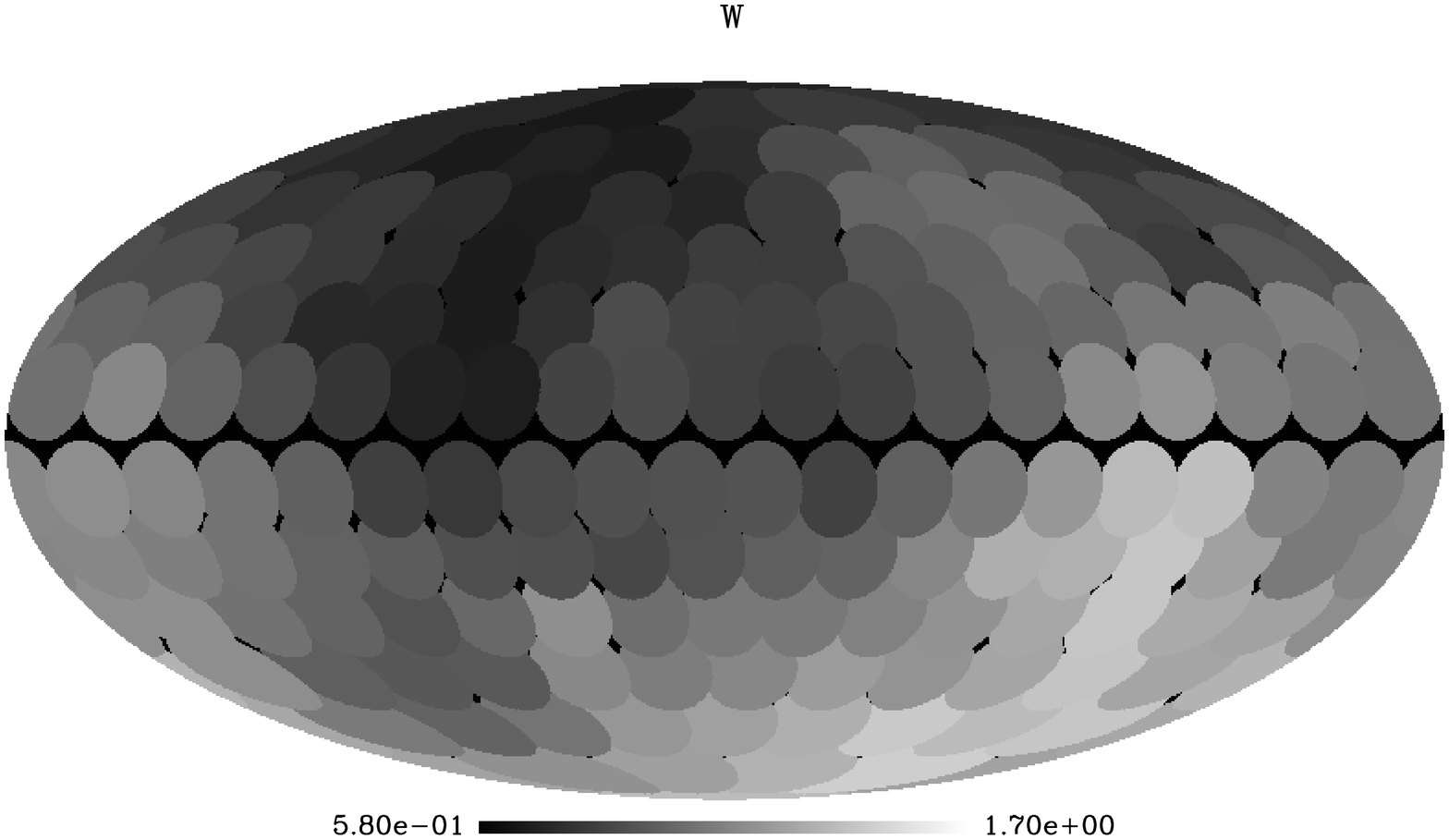,angle=90,height=3.cm,width=6.cm}\hspace*{2cm}
\psfig{figure=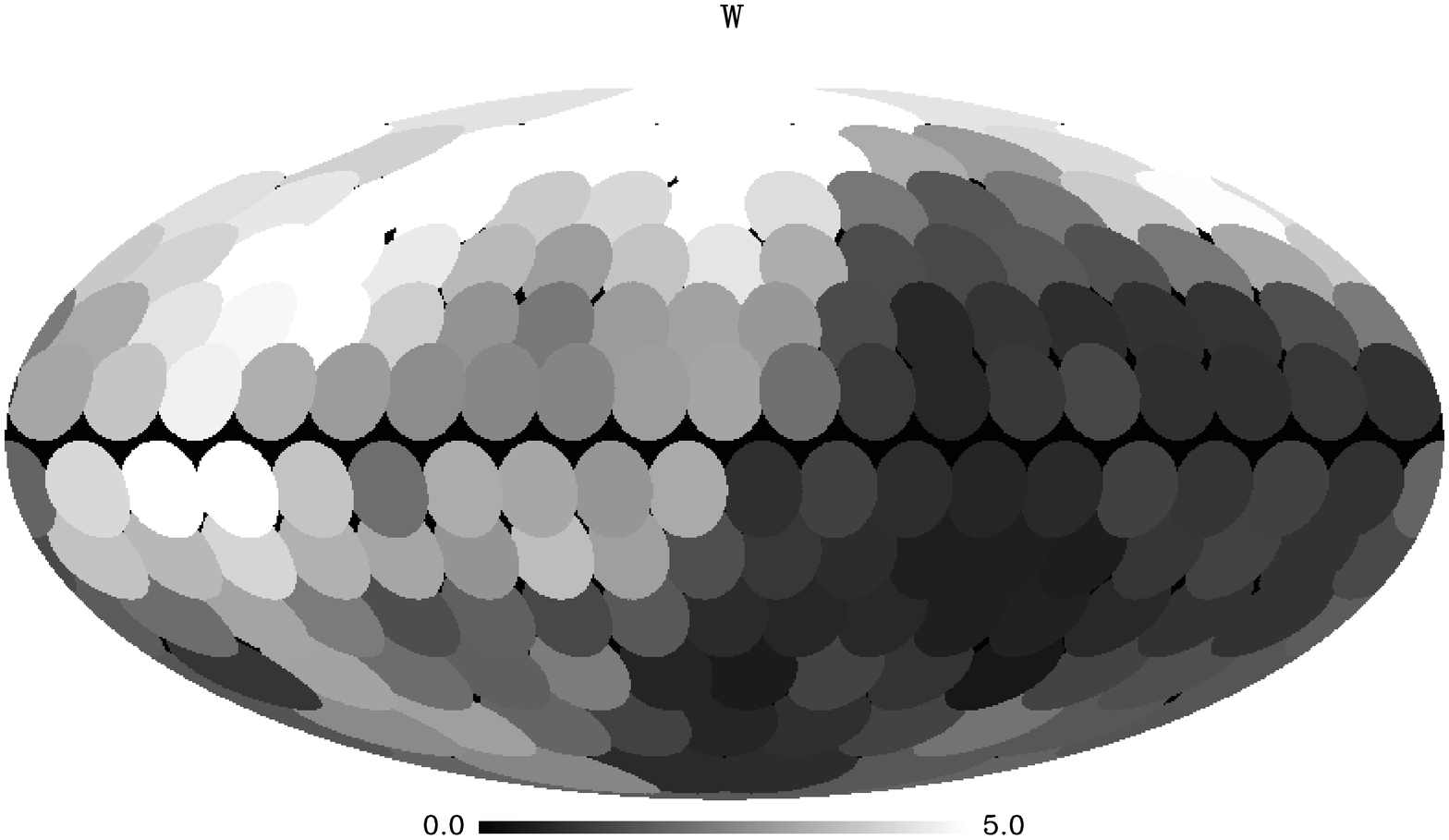,angle=90,height=3.cm,width=6.cm}
\caption{The discs in the figures show the centres of hemispheres on which the power spectrum (left column) and local curvature properties (right column) has been estimated. For the power spectrum, the ratio of large scale power ($\ell=5-20$) between opposite hemispheres is shown (dark discs show a low ratio), whereas for the local curvature the reduced $\chi^2$ of the lake function for the given hemisphere is shown (bright discs show high $\chi^2$, thus non-Gaussianity). Two \emph{WMAP} frequency channels are shown, Q (upper row) and W (lower row).
\label{fig:maps}}
\end{figure}

Similarly to \cite{dore} the points of the renormalised map $T(\theta,\phi)\rightarrow (T(\theta,\phi)-\VEV{T})/\sigma(T)$ can be classified in three types; {\it hills}, {\it lakes} and {\it saddles} where the eigenvalues of the Hessian are both positive, both negative or of opposite sign respectively. One can hence probe non-Gaussianity by evaluating the proportions of hills, lakes and saddles above a certain threshold $\nu$ in the normalised (and smoothed) map on a given part of the sky (see \cite{dore}). The Hessian on the sphere is calculated using the covariant second derivatives \cite{eriksen2,curvat,schmalzing} evaluated in spherical harmonic space. More details can be found in \cite{paolo}. To find the degree of non-Gaussianity, a $\chi^2$ procedure was used. This procedure was applied to the \emph{WMAP} data using the Kp2 sky cut. In \cite{curvat} it was shown that the northern hemisphere was non-Gaussian at the $2-3\sigma$ level for smoothing scales between $1^\circ-5^\circ$ in agreement with similar results using the Minkowski functionals \cite{eriksen2}. No significant non-Gaussianity was found when applying the test on the southern hemisphere or on the full sky. In figure \ref{fig:maps} I have shown this $\chi^2$ for the lake counts, using a $5^\circ$ smoothing scale (the discs show the centres of the hemispheres on which the test has been performed). One can see that the non-Gaussian (high $\chi^2$) region, corresponds well to the region where the large scale power spectrum is low.
\section{Discussion and conclusion}

The CMB power spectrum at large scales ($\ell=2-40$) is found to be significantly much lower in the northern hemisphere (ecliptic and galactic frame of reference) than in the southern \cite{p3,eriksen1}, giving a different estimate of the cosmological parameters in these hemispheres \cite{p5}. Moreover, testing the local curvature properties, a highly non-Gaussian region between the north galactic and north ecliptic pole is found. This is similar to the region where the power spectrum is found to be particularly low. As the curvature test is power spectrum dependent, the non-Gaussian signature could arise due to the difference in the spectrum rather than a non-Gaussian distribution. I have performed the test of local curvature assuming the best fit spectrum found on the northern hemisphere of maximum asymmetry \cite{p5}. In this case, the non-Gaussian detection in the northern hemisphere is still significant ($2\sigma$ level) but less significant than when using the best fit full sky spectrum. Furthermore, the $\chi^2$ in the southern hemisphere becomes particularly low with respect to simulations. Thus, the non-Gaussian detection does not appear to be only due to the difference in the power spectrum between the two hemispheres.

There have been detections of foregrounds residuals in the \emph{WMAP} data \cite{p3,nas1,nas2,nas3,nas4,eriksen3,schwarz,slosar}. Foregrounds have a well known frequency dependence and incorrectly subtracted foregrounds would be expected to show up at a different level in different frequency channels. The asymmetry however, has been seen at a similar level in the different \emph{WMAP} frequency bands as well as in the \emph{ILC} map \cite{p3} which was constructed using a different foreground subtraction method. A similar asymmetric distribution of the power spectrum was also found in the \emph{COBE} data \cite{eriksen1}. The fact that the \emph{COBE}-DMR data is susceptible to different parasitic signals compared to \emph{WMAP} argues against an explanation for the results in terms of a systematic effect. 

Nevertheless, if some unknown systematic effect or foreground residual in some unknown way gives rise to the asymmetry, the above mentioned results do not give a clear answer to {\it where} the contamination is found. There have been detections of non-Gaussianity on the northern hemisphere using local curvature and Minkowski functionals \cite{park,eriksen2,curvat}. However for the three-point function, the southern hemisphere shows a non-Gaussian signal whereas the northern hemisphere is particularly Gaussian \cite{eriksen1}. Also using wavelets \cite{vielva,cruz}, a non-Gaussian feature in the southern hemisphere was detected. An analysis of the cosmological parameters on the different hemispheres \cite{p5} reveals a very high value of the optical depth $\tau\sim0.23$ in the south. 

I have shown that the asymmetry in the power spectrum may be related to the non-Gaussian detection found in one part of the sphere by the local curvature test. Other methods have revealed a similar asymmetry\cite{park,eriksen1,vielva,eriksen2,copi,curvat,wandelt,cruz,land} which may have the same origin because of a similar direction and frequency independence. To clarify whether the asymmetry can be explained in terms of foreground residuals, systematic effects or possibly new physics, the 2 year \emph{WMAP} data and ultimately the forthcoming Planck satellite mission will be needed.
\section*{Acknowledgements}
This work was done in collaboration with A. J. Banday and K. M. G\'orski (for the power spectrum/parameter estimation \cite{p3,p5}) and A. Balbi (parameter estimation \cite{p5}). For the curvature test \cite{curvat}, my collaborators were P. Cabella, D. Marinucci and N. Vittorio. I acknowledge financial support from the CMBNet Research Training Network.

\end{document}